\documentclass[manuscript]{rmaa}
\usepackage{rmaacite}
\newcommand\msun{\mbox{M$_\odot$}}
\newcommand\I{{\'\i}}
\newcommand\dydo{$\Delta Y/\Delta O$}
\newcommand\dydz{$\Delta Y/\Delta Z$}
\newcommand\HII{H\,{\sc ii} }

\title{CARBON AND OXYGEN GALACTIC ABUNDANCE GRADIENTS:
A Comparative Study of Stellar Yields}

\author{L. Carigi 
\affil{Instituto de Astronom\'{\i}a, UNAM, M\'exico}}

\fulladdresses{
\item L. Carigi: Instituto de Astronom\'\i a,
  UNAM, M\'exico D.F., 
   M\'exico (carigi@astroscu.unam.mx).}

\shortauthor{Carigi}
\shorttitle{C and O Galactic Abundance Gradients}

\abstract{

Chemical evolution models for the Galactic disk under an inside-out formation scenario
are presented for seven sets of stellar yields, all of them
metal dependent. In particular,
the effects of yields from massive and low-and-intermediate-mass stars on 
the C/O chemical history of the solar neighborhood 
and on abundance gradients are discussed comparing the
predictions with abundance ratios from nearby
\HII regions, B-stars, dwarf stars, and the Sun. 

In the solar vicinity, the increase of C/O with metallicity is due to massive stars alone.
Models with yields by Maeder (1992) or by Portinari, Chiosi, \& Bressan (1998) reproduce
the observed  C/O increase with metallicity, while models assuming yields
by Woosley \& Weaver (1995) and Woosley, Langer, \& Weaver (1993) do not.
Models based on yields by Maeder (1992) are in agreement with the C/O gradient, while 
those based on yields by Portinari et al. (1998) or by
Woosley \& Weaver (1995) and Woosley et al. (1993) are not. To match both, the C/O enrichment history
of the solar neighborhood and the C/O local gradient, models call for: i) yields that take into account
metal-dependent stellar winds; and ii) for massive stars, a mass-loss rate with a metallicity dependence
not as simple as $Z^{0.5}$.

C/O abundance ratios predicted with yields by Marigo, Bressan, \& Chiosi (1996, 1998) are higher than 
those with yields by
Renzini \& Voli (1981), which in turn are higher than those with yields  by the van den Hoek \& Groenewegen (1997).

The observed O/H and surface density distributions are matched using any yield set.
\dydz\ values are computed for different combinations of seven sets of stellar
yields. All models agree with the \dydz\ value derived from M17, but are smaller than
those derived from other \HII regions.}

\begin{document}
\maketitle
\section{INTRODUCTION}

In a chemical evolution model that assumes an infall of H and He gas without mass
loss by galactic winds or radial flows,
the predicted C/O abundance ratio depends mainly on the stellar yields and on the
initial mass function. By fixing the initial mass function, this work explores the
use of C/O to test stellar yields.

The stellar yield of an element $X_i$, being the mass fraction of a star of initial mass $m$ 
converted to $X_i$ and ejected to the interstellar medium (ISM), strongly depends
on the assumptions of the stellar evolution modeling. 
By the 80', no stellar yield set for massive stars ($m > 8$ \msun) of different
metallicities had been computed. In this decade, Maeder (1992, M92) published
He, C, O, and $Z$ stellar yields for two initial metallicities 
($Z=0.001$ and $0.02$) for objects with initial mass in the 9-120 \msun \ range. Later on,
Woosley \& Weaver (1995, WW) calculated stellar yields for 80 elements and isotopes
for stars with initial masses between 11 and 40 \msun \ and 
with five initial metallicities ($Z/Z_{\odot}=$ 0, $ 10^{-4}$,  0.01, 0.1, and 1),
without initial abundance ratios scaled to solar. More
recently, Portinari, Chiosi, \& Bressan (1998, PCB) published stellar yields of 17 elemental species
for a wide range of metallicities ($Z=0.0004$, 0.004, 0.008, 0.02, and 0.05) and masses (6-120 \msun).

For low and intermediate mass stars
(LIMS, $0.8 < m/{\rm M}_\odot < 8$), stellar yields have been calculated by Renzini \& Voli (1981, RV),
van den Hoek \& Groenewegen (1997, HG), and Marigo, Bressan, \& Chiosi (1996, 1998, MBC).
While HG produce a grid of stellar yields for five metallicities
($Z=$0.001, 0.004, 0.008, 0.02, and 0.04), RV and MBC sampled yields for only a pair of metallicities
($Z=$0.004, 0.02 and $Z=$0.008, 0.02 respectively). A combination of PCB and MBC provides the
only complete grid of stellar yields computed from the same evolutionary tracks (Padova group). 

Carigi (1994) has shown that the evolution of [C/O] with [O/H] in the solar vicinity can be explained
by M92 yields due to the dependence of the C and O yields on initial $Z$.
Prantzos, Vangioni-Flam, \& Chauveau (1994) concluded that the metallicity
dependent M92 yields are also able to reproduce the growth of [C/O] in the solar neighborhood.
Extending the modeling beyond the solar vicinity, this work follows the
[C/O] abundance and evolution with chemical evolution models for the galactic disk, with the aim to study
the behavior of differences of sets of metal-dependent yields from both, massive stars and LIMS.

In general, chemical evolution models of the solar vicinity and the Galactic disk assume a given yield
set, either M92 yields (e.g. Giovagnoli \& Tosi 1995, Prantzos \& Aubert 1995, Carigi 1996),
WW yields  (e.g. Timmes, Woosley, \& Weaver 1995, Chiappini, Matteucci \& Gratton 1997,
Allen, Carigi, \& Peimbert 1998, Prantzos \& Silk 1998), or Padova yields (e.g. Portinari et al. 1998, 
Tantalo et al. 1998).
Since these yield sets have substantial differences, it is not straightforward to separate the effects of
yields from other model assumptions when intercomparing the different studies. In this work,
consistent chemical evolution models are build to find out 
which yield sets reproduce better the ISM abundances. The predicted abundances from
the different sets of stellar yields are compared for the four elements in common among
all sets (H, He, C, and O).

In this paper, all chemical abundances are presented by number, with the exception of \dydo, and \dydz,
that are given by mass.

The paper has been organized as follows:
The observational constraints on the models are presented in \S 2.
\S 3 describes the assumptions adopted in the chemical evolution models.
The model predictions are then shown and briefly discussed in
in \S 4. The global discussion and conclusions are presented in \S 5 and \S 6, respectively.

\section {OBSERVATIONAL CONSTRAINTS}

 In this work, the compilation by Peimbert (1999) and the recent data of Esteban et al.
(1998, 1999a, 1999b) are used as abundance constraints for the models. 
He/H, C/H, and O/H values, for the Galactic \HII regions M17, M8, and Orion are taken from
Peimbert's Table 1, but corrected for dust 
depletion, adding 0.08 dex and 0.10 dex to the O and C abundances respectively,
as suggested by Esteban et al. (1998). In the modeling, I attempt to fit the abundance
gradients to be within the average values by Peimbert (1999) and 
those obtained from the observations by Esteban et al. (1998, 1999a, 1999b),
considering temperature fluctuations  ($t^2 > 0.00$). 
Average O/H gradients computed by Peimbert (1999) are based on the gradients from
Esteban et al. (1998, 1999a, 1999b), Shaver et al. (1983), and Deharveng et al. (1999).
Since in the literature there are C/O values based on recombination lines only
for M17, M8, and Orion, the average C/H gradient is obtained from
Esteban et al. (1998, 1999a, 1999b) and Peimbert et al. (1992).
The relative enrichment
\dydo \ and \dydz \ values for  M17, M8, and Orion are taken from Peimbert (1999, $t^2 > 0.00$), which
have already been corrected for dust (also as suggested by Esteban et al. 1998).

 O/H abundances and gradient from B-stars are in agreement with those from \HII regions
(Gummersbach et  al. 1998, Smartt \& Rolleston 1997), but the C/H values are lower by 
at least 0.3 dex, and the C/H gradient is twice as flat (Gummersbach et  al. 1998). 
The gradients from B-stars shown in Table 3 were computed using the data of Gummersbach et al. (1998)
and Smartt \& Rolleston (1997) for the galactocentric range considered here ($4 < r < 10$ kpc).
The innermost B-star of both data sets was eliminated from the fit, given their uncertain abundances as
discussed by the respective authors.
The He/H data by Gummersbach et al. (1998) are not considered in this work given that the 
large dispersion in abundances ($-1.1 <{\rm log(He/H)} < -0.5 $, corresponding to $14 < $ \dydo $ < 150$)
does not provide a tight enough constraint for the chemical evolution modeling.
The large differences in He abundance observed among B-stars may come from a
contamination in the outer stellar layers by helium produced by the stars, combined with errors and
other uncertainties in the abundance determinations.
Although B-stars and \HII regions show a different slope of the C/O gradient, both of them are negative.
In this study, the presence (not the slope magnitude) of a negative C/O gradient in the galactic disk is used as
another observational constraint.

The model C/O history for the solar vicinity is constrained with C/O abundances from
dwarf stars located 1 kpc around the Sun (Gustafsson et al. 1999)
and the solar value from Grevesse \& Sauval (1998).
Since C/H values from B-stars are lower than those from \HII regions, two different
present-day C/O values are presented.
The increase of C/O with metallicity in the solar vicinity provides yet another observational
constraint.

Planetary nebulae abundances are not be used here as constraints because C is produced
by their progenitors. The models by Renzini \& Voli (1981), van den Hoek \& Groenewegen (1997), and
Marigo et al. (1996, 1998) predict C enrichment in the envelope during the
evolution of stars with masses lower than $\sim$ 8 \msun. Moreover, Peimbert, 
Torres-Peimbert, \& Luridiana (1995) and Peimbert, Luridiana, \& Torres-Peimbert (1995)
found C/H values higher than the solar value, by at least 0.1 dex,
for the vast majority of planetary nebulae in their sample. 

 Another important observational constraint is gas consumption,
measured as the ratio of gas to total surface mass densities,
$\sigma_{gas}/\sigma_{tot}$. In this work the observed $\sigma_{gas}$ distribution, as
compiled by Matteucci \& Chiappini (1999), is used together with an exponential $\sigma_{tot}$
distribution with a 3 kpc scale-length and a 
$\sigma_{tot}(r_\odot)=45\ M_\odot pc^{-2}$ amplitude (Kuijken \& Gilmore 1991).
A galactocentric distant for the Sun, $r_\odot$, of 8 kpc is adopted here.

Summarizing, the just-described observational constraints are used in this study as follows:
 a) All models (for each and every yield set) are build to exactly reproduce:
    i) the observed gas fraction distribution of the galaxy, $\sigma_{gas}/\sigma_{tot}$; and
    ii) the observed O/H galactic gradient.
 b) In order to study the differences among available yields sets, 
the model predictions are then compared to:
    i) the observed rise of C/O with metallicity 
(or equivalently with time) in the solar neighborhood; and
    ii) the observed decrease of the C/O abundance with galactocentric distance.

\section{ CHEMICAL EVOLUTION MODELING}

 The scope of this work is to explore models assuming the different sets of metal-dependent stellar yields
available in the literature.

 All models are built to reproduce two observational 
constraints: the O/H abundance gradient from \HII regions and B-stars, 
and the $\sigma_{gas}$ distribution from 4 to 10 kpc.
Models are very similar to the infall model of Allen, Carigi, \& Peimbert (1998), but
adopting somewhat different assumptions:

 a) The star formation rate is set proportional to a power of $\sigma_{gas}$ and
   $\sigma_{tot}$: $ SFR(r,t) = \nu \ \sigma^x_{gas}(r,t) \ \sigma^{x-1}_{tot}(r,t)$, where
   $\nu$ and $x$ are constant in time and space, and fixed in such away that the
   observational constraints are reached after 13 Gyr, the age of the model.

 b) Three sets of metal-dependent stellar yields from massive stars are used: 
   i) Geneva yields: M92  for $9 \leq m/\msun \leq 85$ (high mass-loss rate);
   ii) Santa Cruz yields: WW 
   for $11 \leq m/\msun \leq 40$ (models ``B" for 30, 35 and 40 \msun),
   and WLW for $ m = 60 \ \msun$ and $m=85 \ \msun$;
Since $Z_{max}=Z_{\odot}$ for the Geneva and Santa Cruz groups, then yields (${Z>Z_\odot}$) = yields($Z_\odot$).
   iii) PCB yields: Portinari et al.  (1998) for $9 \leq m/\msun \leq 85$.
In all cases, effects due to black hole formation are not considered.

 c) Three sets of metal-dependent stellar yields for LIMS are used:
    i) RV yields: Renzini \& Voli (1981) for $1 \leq m/\msun \leq 3$ ($\alpha=0.0$,$\eta=0.3$, case A),
        for $3 \leq m/\msun \leq 8$ ($\alpha=1.5$, $\eta=0.3$, case A).
    ii) HG yields: van den Hoek \& Groenewegen (1997) for $0.8 \leq m/\msun \leq 8$ ($\eta_{AGB}=0.4$,$M_{HBB}=0.8$). 
    iii) MBCP yields: Marigo et al. (1996) for $0.8 \leq m/\msun \leq 3$; 
        Marigo et al. (1998) for 4, 4.5, and 5 \msun;
	Portinari et al.  (1998) for 6 and 7 \msun.
  
 d) The fraction of binary systems predecessors of  SNIa is taken as $A=0.07$.
   Contrary to Carigi (1994), SNIb produced by binary systems are now not considered.

Carigi (1996) and Allen, Carigi, \& Peimbert (1998)  successfully
modeled galactic abundance gradients
considering the IMF by Kroupa, Tout, \& Gilmore (1993). Therefore, in
this work 
the same IMF is assumed for a 0.01 to 85 \msun mass range.
Moreover, this IMF is maintained constant in time and space, based on the
conclusion of 
Chiappini, Matteucci, \& Padoan (2000) that a constant IMF is still the
best assumption
to explain the observational constraints in the Milky Way.

In the present exercise, to avoid spurious mass extrapolation
effects for a fair comparison of the predictions
from the different yield sets, the
stellar mass range was limited so as to be lower than 85 \msun (the maximum
mass common to the three sets from massive stars).

Since the code does not use the instantaneous recycling approximation,
it is necessary to estimate, at
each time step, the stellar yields of all dying stars at the
metallicities at their birth times.
Therefore, the stellar yields from the different groups have been
interpolated in both mass and metallicity,
taking in consideration their different samplings. In particular, given
the number of initial metallicities
considered by stellar evolution models, and the yields $Z$-dependencies,
I have assumed:
a) a linear interpolation with mass for yields of massive stars and
LIMS;
b) a linear interpolation with metallicity for the Santa Cruz, PCB, RV,
HG and MBCP yields; and
c) three kinds of $Z$ interpolations (linear, as $Z^{0.5}$, and $Z^2$)
for the Geneva yields.
It is important to remark that the HG yields vary almost linearly with
$Z$,
but the $Z$-behavior of yields of massive stars is not well determined:
the PCB yields dependency on $Z$ is not simple and the Santa Cruz yields
are almost independent of $Z$.

 Input parameters for the set of models are summarized in Table 1.
Column (2) lists $x$, the SFR power on $\sigma_{gas}$, and column (3) the efficiency factor,
$\nu$, of each model. The slope of the final abundance gradients is basically governed by $x$;
while $\nu$ and $x$ together determine the O abundance. 
Models are divided in three groups (column 4), assuming M92, WW\&WLW, or PCB yields.
M92 and WW\&WLW yields have been complemented with RV, HG, and MBCP yields.
Three subgroups of models have been computed according to the type of interpolation used ($Z^n$, $n=$0.5, 1, or 2) between the two metallicities 
considered by Maeder (1992).
The table lists just one model for the PCB yields (C-model), since
the Padova yields (PCB+MBCP) form a complete and consistent set.

\begin{table}
 \caption{Model Input Parameters}
  \begin{center}
   \begin{tabular}{lcccc}\hline\hline
{Model} & {$x$ \ $^{a}$}  &
{$\nu$} (Gyr$^{-1}$ ($M_\odot$ pc$^{-2})^{2(1-x)}$)  $^{a}$ &
{Massive Stars + LIMS Yields}  $^{b}$  & {$n$}  $^{c}$ \\ \hline
MX0 & 1.22 & 0.048 & M92 + X & 0.5 \\
MX1 & 1.22 & 0.040 & M92 + X & 1.0 \\
MX2 & 1.22 & 0.033 & M92 + X & 2.0 \\
 & & & \\
WX & 1.13 & 0.075 & WW\&WLW + X & \\
 & & & \\
C & 1.15 & 0.045 & PCB + MBC & \\ \hline\hline
    \end{tabular}
   \label{tab:1}
 \end{center}
$^{a}$  SFR $ = \nu \ \sigma^x_{gas} \ \sigma^{x-1}_{tot}$

$^{b}$  X = RV, HG, MBCP

$^{c}$  Interpolation type ($Z^n$)
\end{table}

Since LIMS do not produce O,
 each set of yields from massive stars requires different values of $x$ and $\nu$ to match the same
observational constraints. Since the M92-O yields for $m > 25$ \msun \ decrease with metallicity,
 O gradients saturate quicker, calling for higher $x$ values. On the other hand,
WW-O yields are almost independent of metallicity, and WLW-O yields 
are lower than M92-O yields at low metallicity, 
forcing higher $\nu$ values
for the WW\&WLW modeling to reach the observed O/H of \HII regions.
Since PCB-O yields are lower than M92-yields at low $Z$,
it is required that
$\nu_{\rm {C-model}} \sim \nu_{\rm {M-models}}$ when the interpolation is weighted to high metallicities. 

Since the yields for supermetallic stars are similar to those by WW, 
then $x_{\rm {C-model}} \sim \nu_{\rm {W-models}}$.

Since infall models of the present kind produce abundance gradients that flatten out with time (Carigi 1996),
the M-models require higher $x$ exponents to reproduce the O/H gradient at a fixed age (13 Gyr).

\section{MODEL RESULTS}    

\begin{figure}
\begin{center}
    \leavevmode
    \includegraphics[width=\textwidth]{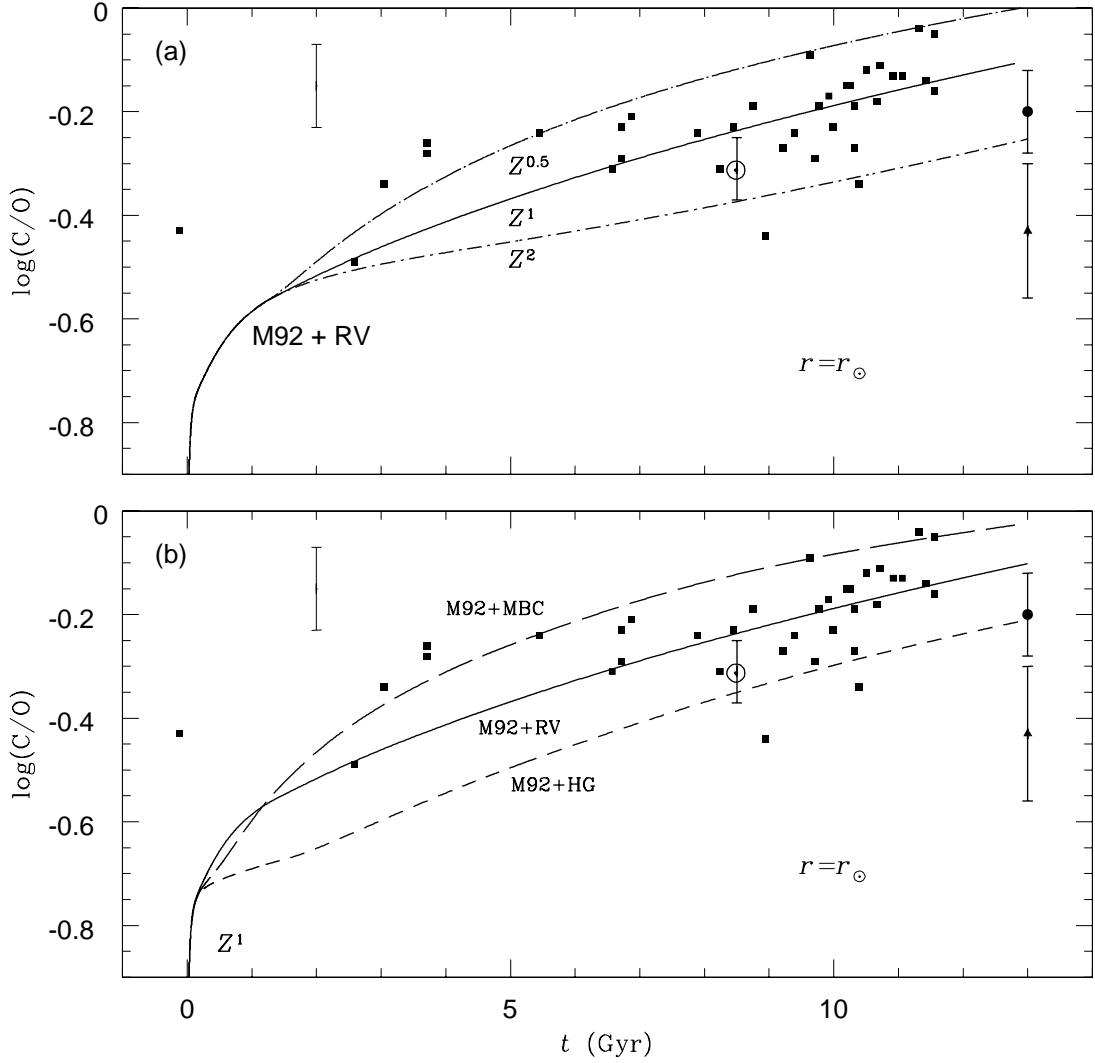}
    \caption{ \footnotesize{
The C/O evolution of the solar vicinity.
Model predictions are presented as follows:
(a) Considering three different interpolations ($Z^{0.5}$, $Z^1$, and $Z^2$) of the massive-stars yields by Maeder (1992) and
LIMS yields by Renzini \& Voli (1981).
{\it Long-dash-dot lines:} MRV0-model,
{\it continuous lines:} MRV1-model,
{\it short-dash-dot lines:} MRV2-model.
(b) Assuming the $Z^1$ interpolation  for yields by Maeder (1992) and three set of LIMS yields by:
Marigo et al. (1996, 1998), Renzini \& Voli (1981), or van den Hoek \& Groenewegen (1997).
 {\it Long-dash lines:} MMBC1-model,
{\it continuous lines:} MRV1-model,
{\it short-dash lines:} MHG1-model.
Observational data are as follows:
{\it filled circle:} computed value at $r=r_\odot$ from radial gradients by Peimbert (1999);
{\it filled triangle:} average value for the five B-stars at $r=r_\odot+0.4$ kpc
from data by Gummersbach et  al. (1998);
{\it filled squares:} dwarf stars at $r=r_\odot \pm 1$ kpc from Gustafsson et al. (1999);
{$\odot$:} solar value from Grevesse \& Sauval (1998).
The ages of the dwarf stars were scaled to the age of the models.}}
\label{fig:f1}
\end{center}
\end{figure}

All models match well several observational constraints at the
solar vicinity, e.g. age-metallicity relation, G-dwarf distribution,
[C/Fe] and
[O/Fe] vs [Fe/H] evolution. For a SFR$ = \nu \ \sigma^x_{gas} \
\sigma^{x-1}_{tot}$ law, the present models adopt $x$ in the range 1.13
and
1.22. Predictions for the solar vicinity do not change significantly when
the $x$ exponent varies from 1 to 2 (Carigi 1996). This work
does not then discuss how well the observed relations are matched, since the
predicted
relations are very similar to those of the best model of Carigi (1996),
where her Fig 4 already summarizes the comparison with the observational
data.

\subsection{C/O evolution of the solar vicinity assuming different yields}

Figure 1 shows the C/O evolution in the solar vicinity
as predicted by models that consider yields by Maeder (1992).
Since the dependence of these yields for massive stars with $Z$ is not well known,
the influence of the assumed type of interpolation on
on the C/O ratio is explored in Figure 1a. 
The figure shows that, between 1.5 and 9 Gyr, the C/O rise in time slows down as the
interpolation exponent $n$ increases ($Z^n$);
out of this lapse, the increase is basically the same for all reasonable $n$s.
It is then important to compute yields for intermediate metallicities to reduce
these interpolation ambiguities. Besides interpolations effects, Figure 1b also
explores the influence of LIMS on the C/O evolution, since these stars are important
carbon producers. After 1 Gyr or so, RV predict less carbon production than MBC
but more than HG.

 The best models from the M92 yields among the different combinations of LIMS yields and
interpolation types assumed (M-models = MRV1, MHG1, MMBC2) are shown in Figure 2.
Results from models assuming the Santa Cruz and Padova yield sets are also presented
in similar diagrams (Figure 3 and Figure 4 respectively). These results, from the
modeling assuming the different yield sets, are discussed in the next two subsections,
looking first at the C/O evolution in the solar vicinity, and later on at the local
chemical distributions along the galactic disk.

\begin{figure}
    \begin{center}
    \leavevmode
    \includegraphics[width=\textwidth]{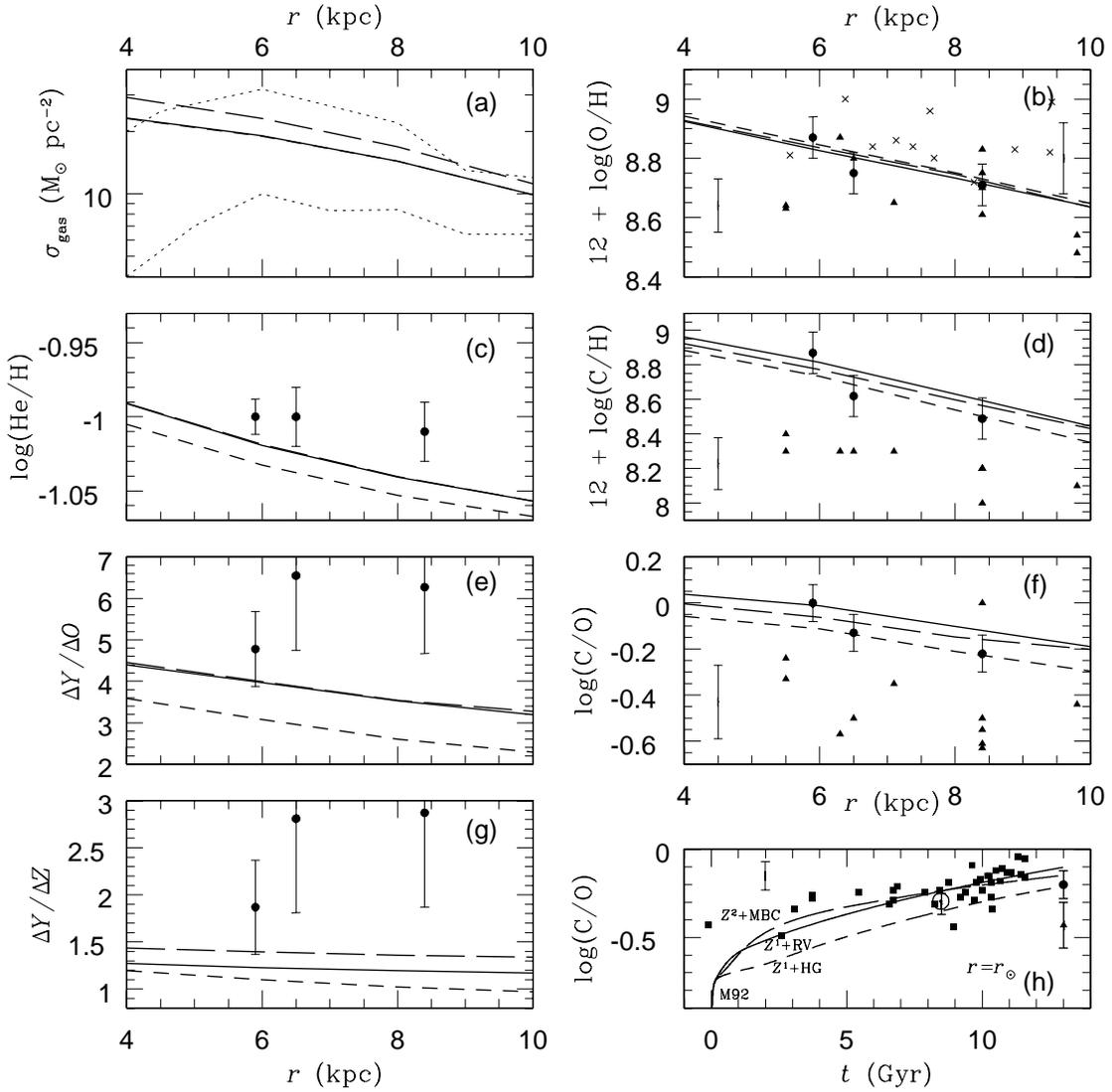}
    \caption{\footnotesize {a) Present-day radial distribution of gas surface mass density,
the area enclosed by dotted lines indicates the data collected by
Matteucci \& Chiappini (1999);
(b)-(g) observed and predicted abundance ratios.
Predictions from the best models considering yields by Maeder (1992):
{\it long-dashed lines:} MMBC2-model,
{\it continuous lines:}  MRV1-model,
{\it short-dash lines:} MHG1-model.
Observational data:
{\it filled circles:} gas and dust in \HII regions from
Peimbert (1999) and Esteban et al. (1998), respectively;
{\it filled triangles:} B-stars from Gummersbach et  al. (1998); and
{\it crosses:} B-stars from  Smartt \& Rolleston (1997).
The observed values correspond to M17, M8, and Orion at adopted galactocentric distances
of 5.9, 6.5, and 8.4 kpc, respectively.
Error bars at the left and the right represent the typical errors from data by
Gummersbach et  al. (1998) and Smartt \& Rolleston (1997), respectively.
(h) The C/O evolution of the solar vicinity, observational data as in Fig 1.
}}
\label{fig:f2}
\end{center}
\end{figure}

 In particular, the (h) panel of each figure presents the predicted C/O chemical enrichment history
of the solar neighborhood, as compared to the following observational data:
i) the observed  abundance in the Sun and in the neighboring disk dwarf stars,
ii) a C/O value at  $r_\odot$ derived from the observed gradient for \HII
regions,
and iii) a C/O average value for the nearest B-stars from Gummersbach et al. (1998).
From these panels, one can see that the C/O abundance always increases with time in the
M-models and the C-model, in agreement with most of the observations (dwarf stars, Sun, and \HII regions),
but the results of the W-models fail to predict the observed recent rise in C/O.

\begin{figure}
        \begin{center}
    \leavevmode
    \includegraphics[width=\textwidth]{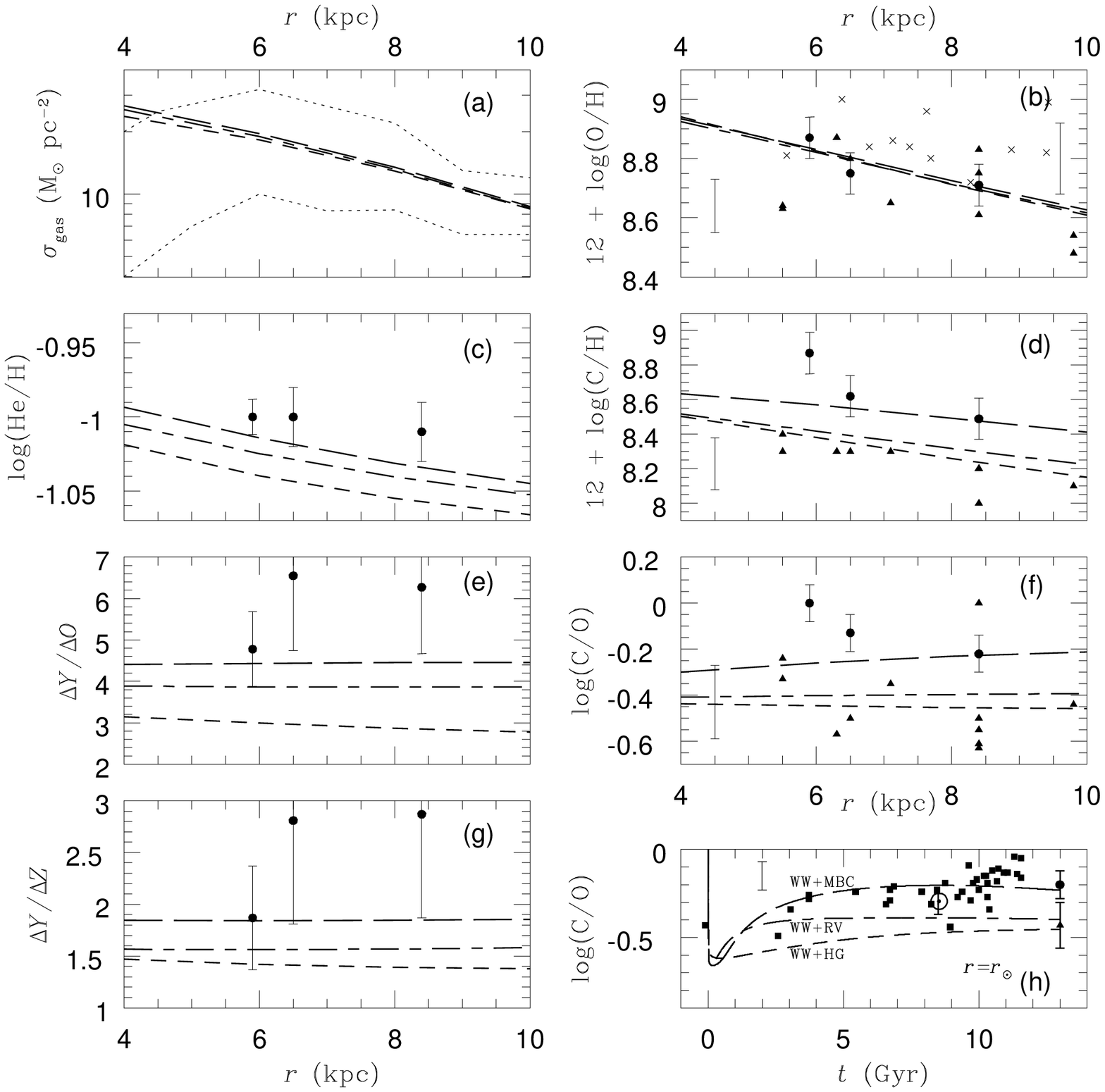}
    \caption{
Same as previous figure but with the
predictions from models assuming yields by
 Woosley \& Weaver (1995)
   and Woosley et al. (1993) for massive stars and
Marigo et al.  (1996, 1998), Renzini \& Voli (1981), or van den Hoek \& Groenewegen (1997)
for low and intermediate mass stars.
{\it Long-dash lines:} WMBC-model,
{\it long-short-dash lines:} WRV-model,
{\it short-dash lines:}  WHG-model.
}
        \label{fig:f3}
        \end{center}
\end{figure}

The discrepancy among the C/O predictions with different yields can be understood. The M92 and PCB-O
yields increase with mass and decrease with metallicity, while the C yields increase with both mass
and metallicity for $Z \leq Z_\odot$. At the onset of the evolution, the WLW yields from very massive
stars ($m > 40 $ \msun) enrich the ISM with ever more oxygen through O-yields
that increase with mass.
 Once these stars die after a few million years,
the C/O of the W-models behaves like in the M-models because, for $m < 40 $ \msun, both sets of
yields are similar at low metallicities. Since the WW yields do not depend
on metallicity,  the W-models reach (at $\sim$ 3.5 Gyr) a plateau in C/O that decreases slowly with time.

\begin{figure}
\begin{center}
    \leavevmode
    \includegraphics[width=\textwidth]{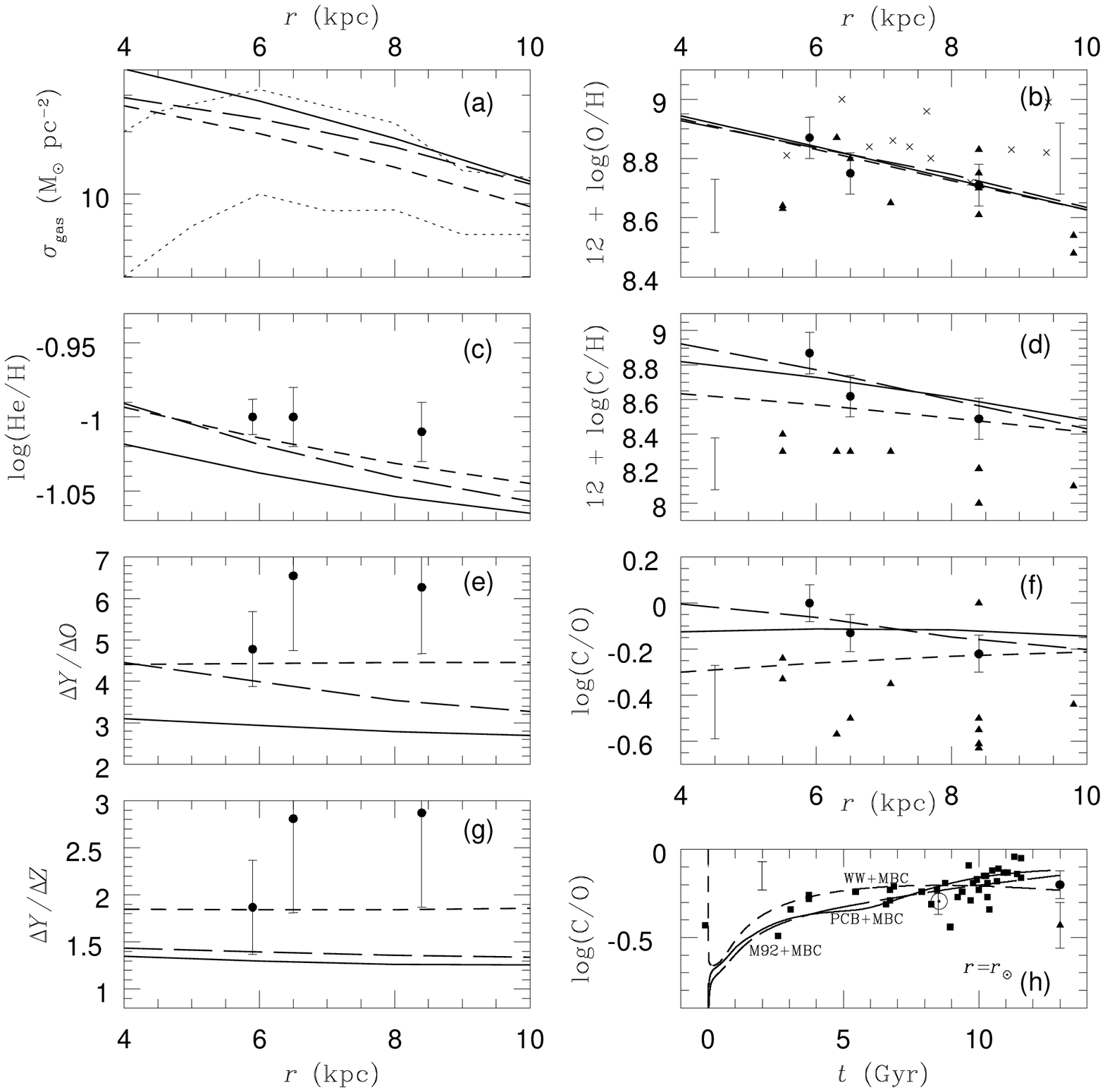}
    \caption{
Predictions from models assuming yields by Marigo et al. (1996, 1998) for low and intermediate mass stars
in combination with yields for massive stars by
and Woosley \& Weaver (1995)
   and Woosley et al. (1993), Portinari et al. (1998), or Maeder (1992).
  {\it Short-dash lines:} WMBC-model,
 {\it continuous lines:} C-model,
 {\it long-dash lines:} MMBC2-model. Observational data points and details as
in Fig. 2.
}
        \label{fig:f4}
        \end{center}
\end{figure}

The C-model predicts two plateaus in the C/O evolution of the solar neighborhood: a first one at $\sim$ 5 Gyr
and another after $\sim$ 11 Gyr.
The later one is due to the contribution of supermetallic stars. These stars have extreme winds that
occur before C and O are synthesized, so their winds are He rich, while C and O are produced
during the SN explosion. These characteristics of the yields from supermetallic stars mimics the
behavior predicted by WW.

 Table 2 summarizes the present-day predictions for the solar neighborhood from all models.
M-models reproduce the observed C/H and C/O values from \HII regions.
WRV and WHG models
predict C/H and C/O values in agreement with B-stars, while the WMBC model results agree with
the C/H and C/O from \HII regions.

\begin{table}
 \caption{Current Predicted and Observed Abundance Ratios for the Solar Vicinity }
  \begin{center}
   \begin{tabular}{lccc}\hline\hline
Model & 12 + log(C/H) & 12+log(O/H) & log(C/O) \\ \hline
MRV0 & 8.74    & 8.74 & +0.00 \\
MRV1 & 8.63    & 8.73 & -0.10 \\
MRV2 & 8.48    & 8.74 & -0.25 \\
MHG0 & 8.66    & 8.75 & -0.10 \\
MHG1 & 8.54    & 8.75 & -0.21 \\
MHG2 & 8.38    & 8.75 & -0.37 \\
MMBC0& 8.81    & 8.75 & +0.06 \\
MMBC1& 8.72    & 8.74 & -0.02 \\
MMBC2& 8.60    & 8.75 & -0.15 \\
WRV & 8.32    & 8.71 & -0.39 \\
WHG & 8.26    & 8.71 & -0.45 \\
WMBC& 8.49    & 8.73 & -0.23 \\
C & 8.62    & 8.73 & -0.12 \\ \hline
Obs. M17, M8, Orion $^{a}$& 8.54 $\pm$ 0.12 & 8.74 $\pm$ 0.07 & -0.20 $\pm$ 0.08 \\
Obs. \HII regions Average $^{a}$       & 8.53 $\pm$ 0.12 & 8.74 $\pm$ 0.07 & -0.21 $\pm$ 0.08 \\
Obs. B-stars $^{b}$  & 8.16 $\pm$ 0.12 & 8.62 $\pm$ 0.08 & -0.46 $\pm$ 0.13   \\
Obs. B-stars $^{c}$  & ...             & 8.84 $\pm$ 0.10 &      ...              \\
\hline\hline
   \end{tabular}
  \label{tab:2}
 \end{center}
$^{a}$ { Computed from data by Peimbert (1999) at $r_\odot$ }

$^{b}$ { Average values for the five B-stars at $r=r_\odot+0.4$ kpc
from data by Gummersbach et  al. (1998)  }

$^{c}$ { Average values at $r=r_\odot \pm 1$ kpc from
data by Smartt \& Rolleston (1997)  }
\end{table}

 Inspecting Figure 1 and Table 2 it can be noticed that
the average C/O for B-stars is lower than for \HII regions and for dwarf stars,
mainly due to the low C/H values observed in B-stars.

\subsection{Chemical radial distributions in the local disk assuming different yield sets}

 Predictions for the Galactic disk are also summarized in the (a) to (g) panels of Figures 2-4,
together with Table 3. From these figures and table it can be seen that:

a) The C/H gradients predicted by the best models with M92 yields are similar to
the observed ones from \HII regions. The gradients from the W-models (considering WW\&WLW yields) or 
the C-model are flatter by about 
0.04 dex kpc$^{-1}$ and are close to the gradient traced by B-stars;

b) C/H abundance ratios from the M, WMBC, and C models  agree with the  \HII regions observations, but those from
WRV and WHG models are lower (by at least 0.1 dex) in closer agreement with the C/H of B-stars;

c) M-models are the only ones that reach the high C/H and C/O values of M17;

d) The negative C/O gradients predicted by the M-models are in very good agreement with the observations. 
The gradients predicted by W-models or C-model, being positive, are a poor match to the observed gradient;

e) The C/O values from W-models agree with those determined from B-stars, 
while the C/O values from M and C models agree with the C/O values determined from \HII regions; 

f) M-models and the WMBC-model reproduce the observed He/H abundances in \HII regions, with the former 
yielding the best agreement;

g) RV yields predict lower C and He than MBC but higher than HG;

h) M-models predict strong negative \dydo \  gradients 
while W-models and C-model yield almost no gradient;

i) Predicted \dydz \ gradients are almost flat for all models;

j) With the exception of the WMBC model, all models predict helium to heavy-elements abundance ratios
in the $1.6 <$ \dydz $<1.0$ range, values that are lower than those observed in M8 and Orion.

The C/O gradients from the W-models and the C-model are almost flat because WW yields are almost independent of $Z$
and the very metal-rich stars eject C and O  in SN explosions and not in winds, both facts imply
a saturation in C/O.

The He/H gradients predicted by the M-models are steeper than by the C-model
because the very-massive and super-metallic stars eject a smaller amount of He than
solar metallicity stars (PCB)

\begin{table}
 \caption{Predicted and Observed Present-day Radial Gradients}
  \begin{center}
   \begin{tabular} {lccc} \hline\hline
{Model}$^{a}$ & {C/H}$^{b}$  & {O/H}$^{b}$ & {C/O}$^{b}$ \\ \hline
MRV0 & -0.078    & -0.049    & -0.029 \\
MRV1 & -0.091    & -0.046    & -0.045 \\
MRV2 & -0.108    & -0.045    & -0.063 \\
MHG0 & -0.082    & -0.049    & -0.033 \\
MHG1 & -0.096    & -0.047    & -0.049 \\
MHG2 & -0.112    & -0.049    & -0.063 \\
MMBC0 & -0.065   & -0.049    & -0.015 \\
MMBC1 & -0.074   & -0.046    & -0.028 \\
MMBC2 & -0.087   & -0.045    & -0.042 \\
WRV & -0.051    & -0.053    & +0.002 \\
WHG & -0.061    & -0.057    & -0.004 \\
WMBC& -0.038    & -0.053    & +0.014 \\
C & -0.056    & -0.054    & -0.002 \\ \hline
Obs. \HII regions Average $^{c}$       & -0.106 & -0.055 & -0.051 \\
Obs. B-stars $^{d}$ & -0.059 & -0.038  & -0.026   \\
Obs. B-stars $^{e}$ &  ...     & -0.017  & ...    \\ \hline \hline
   \end{tabular}
  \label{tab:3}
 \end{center}
$^{a}${ Between 6 and 8 kpc}

$^{b}${ in dex ${\rm kpc}^{-1}$ }

$^{c}${ Peimbert (1999) }

$^{d}${ Computed from data by Gummersbach et  al. (1998) ($r_\odot-2.5$
 kpc $ \leq r \leq r_\odot+1.8$ kpc) }

$^{e}${ Computed from data by Smartt \& Rolleston (1997) ($r_\odot-2.5$
 kpc $ \leq r \leq r_\odot+1.9$ kpc)}
\end{table}

\subsection{C/O vs O/H}

In Figure 5
the predicted evolutions  of C/O versus O/H for the solar neighborhood are presented. 
In addition to Galactic data, 
C/O and O/H values for extragalactic \HII regions are shown.
Objects in the halo and in the disk of the solar neighborhood (cylinder of radio
1 kpc centered in the Sun), and in external galaxies call for an increase of C/O
with O/H enrichment.
Only models that reproduce this increase are shown in the figure, so the
W-models are excluded.
The agreement
for log(O/H) $<$ -3.8 is not so good, with the exception of MHG1. 
Again, here one can notice that C yields for intermediate mass stars given by HG are lower than
those obtained by RV or MBCP.

The log(C/O)-log (O/H) relation predicted by the MHG1-model of this
work (Fig 5) is not surprisingly very similar to the one obtained by Henry,
Edmunds, \& K\"oppen (2000), since their best models, either analytical
or numerical, are constructed also adopting the yields of Maeder (1992)
and van den Hoek \& Groenewegen (1997).

\begin{figure}
       \begin{center}
    \leavevmode
    \includegraphics[width=\textwidth]{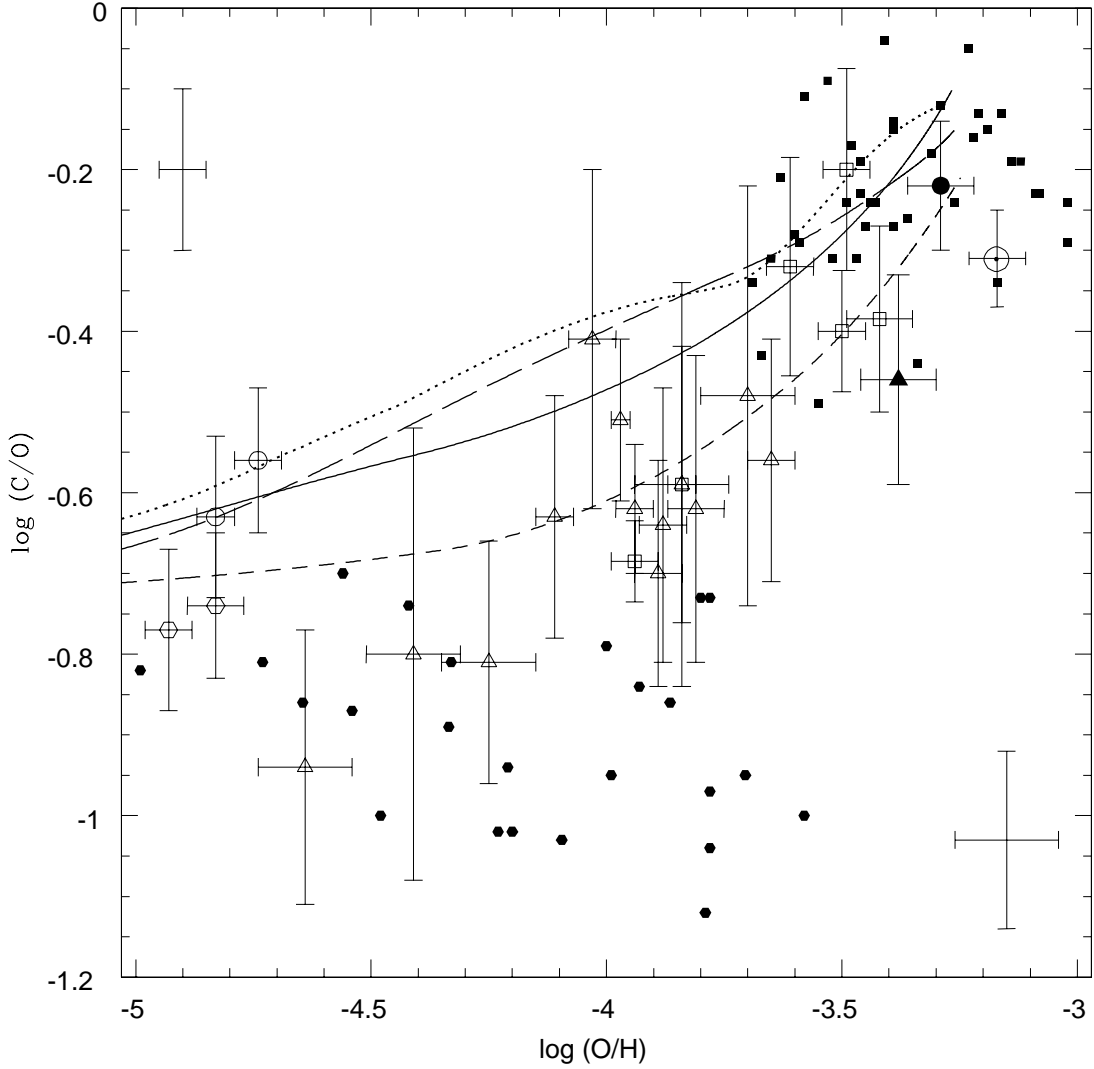}
    \caption{ \scriptsize{
Log(C/O) - log(O/H) relation for $r=r_\odot$.
The predictions of selected models (those that reproduce this relation) are presented as follows:
{\it dotted line:} C-model,
{\it long-dashed line:}  MMBC2-model,
{\it continuous line:}  MRV1-model,
{\it short-dash line:}  MHG1-model.
Observational data for the solar vicinity ({\it filled symbols}) are as follows:
{\it circle}: Orion from Peimbert (1999);
{\it hexagons}: field halo dwarfs from Tomkin et al. (1992);
{\it squares}: dwarf stars  from Gustafsson et al. (1999), as in Fig 2;
{\it triangle}: average B-star
computed from the Gummersbach et al. (1998) data (as in Fig. 2);
{$\odot$}: solar value from Grevesse \& Sauval (1998).
Error bars at the upper-left and the lower-right are typical errors from data by
Gustafsson et al. (1999) and Tomkin et al. (1992), respectively.
Log (O/H) for halo dwarfs is the average of [O/H] values from Tomkin et al. (1992).
Observational data for external galaxies ({\it open symbols}) are as follows:
{\it circles}: I Zw 18NW and I Zw 18SE, values adopted from Figure 4 of Garnett et al. (1999);
{\it hexagons}: I Zw 18NW and I Zw 18SE, abundance ratios taken from Table 5 of Izotov \& Thuan (1999);
abundance data for spiral galaxies ({\it squares}) and
dwarf irregular galaxies and Magellanic Clouds ({\it triangles}) are obtained from Garnett et al. (1999).
Plotted C/O values for spiral galaxies are the average of the abundance ratios for
two different reddening laws.
}}
        \label{fig:f5}
        \end{center}
\end{figure}

A powerful tool for comparing chemical evolution models with interstellar medium
abundances is given by the relation 
\begin{equation}
{\rm log(C/O)}= a \ {\rm log(O/H)} + b.
\end{equation} 
Table 4  presents the $a$ values predicted by the models,
considering different concentric galactic rings:
i) a very local disk, between $r=6$ and $r=8$ kpc, and
ii) the complete area studied in this paper.
Moreover, this table shows  the $a$ values
derived from four sets of observations, two galactic and two extragalactic ones;
for B-stars it was assumed that the C and O abundances correspond to those of
the interstellar medium.
The observed $a$ values indicate that the negative C/O local gradient
may be extended to the whole Galactic disk.
As can be seen from the table, the observed $a$ values are in good agreement with
the M-models and in disagreement with the W-models and the C-model, reiterating the results
presented in Figure 5.
According to the discussion of Figure 5, the best models are:
MRV1, MHG1, and MMBC2.

\begin{table}
 \caption{Present-day Rise across the Galactic Disk of C/O with O/H}
  \begin{center}
   \begin{tabular} {ccccc} \hline\hline
{} & & {Predictions}$^a$ & & {}   \\
{Model} & & {$a_{\Delta r=2}$} &  & {$a_{\Delta r=6}$} \\ \hline
MRV0 &&  0.58   && 0.56  \\
MRV1 &&  0.98   && 0.79  \\
MRV2 &&  1.41   && 1.03  \\
MHG0 && 0.67    && 0.56  \\
MHG1 && 1.04    && 0.81  \\
MHG2 && 1.27    && 1.01  \\
MMBC0 && 0.31   && 0.31  \\
MMBC1 && 0.61   && 0.50  \\
MMBC2 && 0.94   && 0.67  \\
WRV && -0.05    && -0.05 \\
WHG &&  0.07    &&  0.07 \\
WMBC&&  -0.27   && -0.28 \\
C &&  0.04      && 0.06  \\ \hline
& &  Observations$^a$  & &  \\
Object & ${\Delta r}$ (kpc) & & $a_{\Delta r}$ &  \\  \hline
\HII regions Average \ $^b$ & 2.5  && 0.93 $\pm$ 0.60 &\\
B-stars  $^c$               & 12.0 && 1.69 $\pm$ 2.34 &\\
NGC 2403 $^d$               & 5.3  && 0.50 $\pm$ 0.43 &\\
M101     $^d$               & 15.1 && 0.98 $\pm$ 0.29 &\\ \hline \hline
\end{tabular}
\label{tab:4}
\end{center}
{\footnotesize
$^{a}${ The table presents $a$ values derived from $ {\rm log(C/O) =} \ a \ {\rm log(O/H)} + b $, and ${\Delta r}$ is in
 kpc}

$^{b}${ Computed from data by Peimbert (1999) }

$^{c}${ Computed from data by Gummersbach et al. (1998) and Hibbins et al. (1998)}

$^{d}${ Computed from data by Garnett et al. (1999) }
}
\end{table}

\section{DISCUSSION}

 The determinations of C/O from \HII regions in spiral and irregular galaxies (Garnett et al. 1999), 
\HII regions in our Galaxy (Peimbert, Torres-Peimbert \& Ruiz 1992, 
Esteban et al. 1998, 1999a, 1999b), and dwarf stars in the solar vicinity
(Gustafsson et al. 1999); all converge to the conclusion that C/O must increase with metallicity. 
M92 yields can explain this behavior, Padova yields only partially, while WW\&WLW yields not at all.
Other authors (Ferrini, Matteucci, Prantzos, Timmes, and Tosi, see Tosi 1996 for references)
using different chemical evolution models, have predicted flat C/O gradients (Tosi 1996)
because all of them have adopted either yields at a fix $Z$ or
the WW yields, which are almost independent of $Z$. 

Carigi (1994) and Prantzos et al. (1994) had already concluded that the evolution of 
[C/O] with [O/H] or [Fe/H], respectively,  in the solar vicinity
can be explained by the metal dependent yields of Maeder (1992).
The present work have extended these previous ones to the Galactic disk using the recent C 
determinations in \HII regions and B-stars as observational constraints. 
Moreover, seven sets of yields have been considered here, composing the complete sample
of stellar yields dependent on the initial stellar metallicity available in the literature.

 A main difference among these sets of yields resides on the stellar-wind assumptions.
WW do not consider stellar winds at any stage,
and M92 and PCB assume a mass-loss rate proportional to $Z^{0.5}$ and $m^{2.5}$ during 
the post-main-sequence phases.

 C and O yields from M92, PCB, and WW are  similar for $m < 25$ \msun, where stellar
winds are not so relevant. In M92 and PCB, most of the net C is produced before the SN stage,
and since the wind strength increases with metallicity, the C yield increases with $Z$.
On the other hand, O is mainly ejected during the SN stage and, since the amount of O
expelled by the SN depends on the stellar mass just before the SN explosion,
the M92 and PCB-O yield decreases with $Z$.
This fact causes the M-models and the C-model to reproduce the rise of the C/O in the solar
vicinity.

The increase of C yields and the decrease of O yields with $Z$ 
can not be extrapolated to $Z>Z_\odot$.
According to PCB, the C to O yields ratio decreases from $Z=Z_\odot$ to $Z= 2.5 Z_\odot$.
Supermetallic and  massive stars have early and intense winds, which occur before these
stars synthesize C and O. Therefore, their C and O yields are similar to
those of metal-poor stars that mainly eject C and O in the SN explosion.
This behavior of the C and O yields produce a saturation of the C/O abundance ratio at $\sim$ -0.1 dex,
which produces flatter gradients for the C-model.
Since M-models assume that the supermetallic stars behave like metal-solar stars,
the C-model would predict a negative C/O gradient if the mass-loss rate for $Z= 2.5 Z_\odot$
is similar to that for $Z=Z_\odot$ 

 It is important to note that in the W-models the wind contribution to the C and O
yields was not included, since WLW present information to calculate only the He expelled by winds.
If winds were considered in these models, the C yield would certainly be higher but,
given the low weight at the high-mass end of the initial mass function, hardly
enough to reproduce the present-day C/O abundances. When the WLW yields are not taken into
account (by reducing the upper mass-limit to 40 \msun) the predicted C/O 
decreases at times earlier than 0.4 Gyr, but still remains low and basically constant for
the rest of the evolution (Carigi \& Peimbert 2000).

The M-models and C-model agree better with the younger objects in the solar vicinity 
than with the older ones.
The agreement would improve if metal-poor stars eject more oxygen and less carbon
than predicted by a $Z^{0.5}$ law.
The M-models are  in agreement with data for spiral and irregular
galaxies while the W-models are not, as discussed by Garnett et al. (1999).
Then again, the general agreement would
be even better if the C and O yields for metal-poor stars were more dependent on metallicity than
assumed by M92 and PCB. Furthermore,
the C/O value of I Zw 18 determined by Garnett et al. (1995, C/O$=-0.60 \pm 0.10$ dex) would be hard
to explain (not so the lower value of Izotov \& Thuan 1999, $<\rm{log(C/O)}>=-0.78 \pm 0.03>$).
It should be noted that the models presented in Figure 5 were made to reproduce the Galactic disk
and therefore, the comparison of the extragalactic \HII regions and halo objects is only
indicative.  Specific models for each galaxy or for the Galactic halo should be carried out, see for example
the models by Carigi, Col\I n, \& Peimbert (1999) and Carigi \& Peimbert (2000).

 Moreover, the M-models and the C-model agree well with the [C/H], [C/O], and [C/Fe] vs [Fe/H]
relations observed in dwarf stars of the solar vicinity by Gustafsson et al. (1999),
after correcting for the adopted solar abundances (Grevesse \& Sauval 1998 as opposed to
Anders \& Grevesse 1989).
 The predicted relations are matched at high [Fe/H] and become slightly lower than those
observed at lower metallicities; again, a behavior that calls for a stronger metal dependency
of the C and O yields for metal-poor stars than a $Z^{0.5}$ law.
 W-models basically match the relation slopes, but again predict a
lower C abundance than observed  (by $\sim$ 0.3 dex in C/H and $\sim$ 0.2 dex in C/Fe).

Maeder (1993), Woosley \& Weaver (1995) and Portinari et al. (1998) have assumed
different values for the  C$^{12}$($\alpha$,$\gamma$)O$^{16}$ reaction rate.  
It is known that this rate is uncertain by about a factor of 2.
This uncertainty causes minor differences in the behavior of C/O as a function of
 metallicity 
predicted by the models, but more severely affects  the absolute C/O value.
This change in the absolute C/O value does not alter the conclusions of
this paper. In other words, to a very good approximation, a change in the
C$^{12}$($\alpha$,$\gamma$)O$^{16}$ rate modifies $b$ in equation (1), but not $a$.
 
The C/H values derived from  Galactic B-stars are about 0.3 dex lower than those from
Galactic \HII regions and
from  disk dwarf stars of the solar vicinity.
Carbon abundances of B-stars are not easy to understand and may not be representative
of the present-day C in the ISM, because:
i) being so young  ($10^{-3}$ - $10^{-2}$ Gyr), B-stars
would be expected to be richer in C than most dwarf stars, but the observed C/H 
in B-stars in the solar vicinity is very similar to the C of the poorest dwarfs;
ii) the O and N abundances and gradients from B-stars
and from \HII regions are very similar (surprisingly so for N),
so it is difficult to understand why the C/O gradient from B-stars is almost flat,
while the one from \HII regions is not;
iii) a real C/O gradient is supported by the fact that 
other two nearby spiral galaxies, M101 and NGC 2403, also show significant C/O
gradients, -0.04 and -0.05 dex/kpc respectively (Garnett et al. 1999); and
iv) C/O from \HII regions, the Sun, and dwarf stars are consistent with each other,
but not with the C/O from B-stars in the solar vicinity.

It is known that intermediate mass stars ($m < 8$ \msun) produce carbon and not oxygen.
Carigi (1994) found
that yields by Renzini \& Voli (1981) can not explain the C/O increase with the metallicity
in the solar neighborhood. Prantzos et al. (1994) suggested that C-yields higher
that those predicted by Renzini \& Voli could reproduce
the observed rise in C/O. The other two sets of yields for LIMS dependent on $Z$ that exist in the literature 
(van den Hoek \& Groenewegen 1997, and
Marigo et al. 1996, 1998)
can  not explain the recent rise of C/O in the solar vicinity.
However, LIMS have an important role in the early evolution of C/O.

\section{CONCLUSIONS}
 
 Based on basic chemical evolutionary models for the Galactic disk, 
assuming different sets of stellar yields  dependent on metallicity,
the following conclusions are reached:

a) Different sets of stellar yields predict different 
C/O values. The recent rise of C/O with O/H is due mainly to massive stars.
The early rise of C/O with O/H is due to both massive stars and LIMS.

b) Models using stellar yields with stellar winds dependent on $Z$ (Maeder 1992, Portinari et al. 1998) reproduce 
the rise of C/O with time shown by \HII regions, the Sun, and dwarf stars 
within the solar vicinity.

c) Models considering Maeder's yields are also successful in reproducing
the C/O Galactic abundances and gradient determined from \HII regions.

d) The model assuming Padova yields  reproduces C/O abundances but not
C/O Galactic gradients from \HII regions, because stellar winds  $\propto Z^{0.5}$
become too strong for supermetallic stars.

e) Models assuming yields from WW\&WLW reproduce 
only the C/O abundances from B-stars,
but fail to reproduce the other observational constraints.

f) Since the C/H abundances in B-stars are lower than in disk dwarf stars of different ages in 
the solar neighborhood,  while the O/H value is very similar, 
the C/O value from these B-stars may not be a good observational
constraint. 

g) The main difference between the sets of yields of massive stars arises because
Geneva and Padova groups consider stellar winds dependent on metallicity while WW do not,
 and WLW only partially. One can then conclude that $Z$-dependent stellar winds must play an
important role in the chemical enrichment history of the Galaxy.

h) Observations within the solar neighborhood, the Galactic disk, as well as in spiral and 
irregular
galaxies imply that C/O must increase with metallicity. Modeling including not only winds, but
actually metal-dependent winds, is necessary to properly follow the chemical evolution of galaxies.

i) To improve the agreement with the C/O Galactic abundances 
and the C/O evolution with metallicity,
the present models call for a  more complicated mass-loss rate law than
$Z^{0.5}$, assumed by Maeder (1992) and by Portinari et al. (1998):
$\dot{M}_{wind} \propto Z^n $, such that $n> 0.5 $ if $Z \ll Z_\odot$,
$n \sim 0.5$ when $ Z \leq  Z_\odot$ and $0.5 < n \ < 0.7$ for $ Z > Z_\odot$.

j) Models based on yields by Renzini \& Voli (1981) predict less C and He than those
by Marigo et al. (1996, 1998)
and more than those by van den Hoek \& Groenewegen (1997)

k) The \dydz \ value decreases along the sequence of models MBCP - RV - HG.
At the same time, it increases along the PCB -  M92 - WW\&WLW model sequence.

l) The number of \HII regions with known C abundance
is quite small, and the C/O gradient from \HII regions might change with future
C determinations in more \HII regions.
It is important to obtain C abundances from \HII regions located at other galactocentric
distances to determine the Galactic C/O gradient with higher accuracy.

\acknowledgments
I would like to thank Manuel Peimbert for
valuable comments and useful suggestions. 
I also wish to thank Laura Portinari for providing me the massive stars yields of the Padova group.
I am grateful to Don Garnett for supplying me the data for irregular and spiral galaxies.
I also acknowledge the several excellent suggestions by the referee which have improved this final version.
I thank Jes\'us Gonz\'alez for a thorough reading of the manuscript.

\end{document}